\documentclass{elsarticle}
\usepackage{graphicx}
\usepackage{amssymb,amsmath}
\usepackage{stackengine}

\usepackage{doi}
\usepackage[mathlines]{lineno} \modulolinenumbers[5]
\usepackage{xcolor}
\usepackage[nameinlink,noabbrev]{cleveref}
\usepackage{etoolbox} 
\newcommand*\linenomathpatchAMS[1]{%
  \expandafter\pretocmd\csname #1\endcsname {\linenomathAMS}{}{}%
  \expandafter\pretocmd\csname #1*\endcsname{\linenomathAMS}{}{}%
  \expandafter\apptocmd\csname end#1\endcsname {\endlinenomath}{}{}%
  \expandafter\apptocmd\csname end#1*\endcsname{\endlinenomath}{}{}%
}
\expandafter\ifx\linenomath\linenomathWithnumbers
  \let\linenomathAMS\linenomathWithnumbers
  \patchcmd\linenomathAMS{\advance\postdisplaypenalty\linenopenalty}{}{}{}
\else
  \let\linenomathAMS\linenomathNonumbers
\fi
\linenomathpatchAMS{gather}
\linenomathpatchAMS{multline}
\linenomathpatchAMS{align}
\linenomathpatchAMS{alignat}
\linenomathpatchAMS{flalign}
\journal{Chaos, Solitons \& Fractals}
 \newcommand{\ndtot}[3] { \frac{ d^{#3} {#1} } {d #2 ^{#3} }}
 
 \newcommand{\ndpar}[3] { \frac{\partial^{#3} {#1} } {\partial #2 ^{#3} }}
 \newcommand{\OL}[1] {\textrm{\emph{\large O}}\!\left({#1}\right)}
 
\begin{document}

\begin{frontmatter}

\title{Fractal structure of the soliton scattering for the graphene superlattice equation}

\author{Francisca Martin-Vergara\corref{COR}}
 \ead{fmarver@uma.es}
\author{Francisco Rus}
 \ead{fdrus@uma.es}
\author{Francisco R. Villatoro}
 \ead{frvillatoro@uma.es}

\address[eii]{Escuela de Ingenier\'ias Industriales, Dept. de Lenguajes y Ciencias de la Computaci\'on, \\ Universidad de M\'alaga, 29071 M\'alaga, Spain}

\cortext [COR]{Corresponding author: Tel.: +34-951952720; fax: +34-951952542.}

\date{} 

\begin{abstract}
  The graphene superlattice equation, a modified sine-Gordon equation, governs the propagation of solitary electromagnetic waves in a graphene superlattice. This equation has kink solutions without explicit analytical expression, requiring the use of quadrature methods. The inelastic collision of kinks and antikinks with the same but opposite speed is studied numerically for the first time; after their interaction they escape to infinity when its speed is either larger than a critical value or it is inside a series of resonance windows; otherwise, they form a breather-like state that slowly decays by radiating energy. Here, the fractal structure of these resonance windows is characterized by using a multi-index notation and their main features are compared with the predictions of the resonant energy exchange theory showing good agreement. Our results can be interpreted as new evidence in favour of this theory.
\end{abstract}

\begin{keyword}
 Graphene superlattice equation \sep Fractal structure \sep kink-antikink scattering \sep Computational simulations
\end{keyword}

\end{frontmatter}


\section{Introduction}

In 2009, Pavel V. Ratnikov~\cite{Ratnikov2009} introduced the graphene superlattice (GSL) as a sheet of a graphene monoatomic layer deposited on a heterostructure formed by periodically alternating nanometric layers of SiO$_2$ and hBN (hexagonal boron nitride); the layers are arranged such that the hexagonal lattice was exactly under that of the graphene. S. V. Kryuchkov and E. I. Kukhar'~\cite{KryuchkovKukhar2012} showed that the sine-Gordon equation governed the propagation of nonlinear solitary electromagnetic waves in Ratnikov's graphene superlattice subjected to sinusoidal electromagnetic radiation if radiation frequency is much larger than the plasma frequency and the overlap of the wave functions of neighboring wells is very small. The relaxation of the last condition results in a generalized sine-Gordon equation, here on referred to as the graphene superlattice equation~\cite{KryuchkovKukhar2013}; this equation can be either approximated by a double sine-Gordon equation~\cite{KryuchkovChaotic2013} or further generalized to an integro-differential Klein--Gordon equation for electromagnetic waves in the terahertz regime~\cite{KryuchkovEtAl2013}. The study of the solitary wave solutions of the graphene superlattice equation and their interactions could be useful for the development of new technological applications. Note that the main obstacle for the practical application of these superlattices is the strong damping of the solitary waves, which leads to the dynamic chaos in the electrons, as shown by Melnikov method~\cite{KryuchkovChaotic2013,KryuchkovKukhar2015}.

The analysis of the scattering of topological solitons in non-integrable equations was initiated more than forty years ago, but it remains as an active field of research even in 1+1 dimensions~\cite{PhiFour2019,Kevrekidis2019,AskariEtAl2020,ElGanainiKumar2020}.
The graphene superlattice equation (GSLeq) is expected to be non-integrable since it is not in the list of nonlinear Klein--Gordon equations passing the Painlev\'e test~\cite{ClarksonEtAl1986}; hence, its solitary waves are not true solitons, as suggested in Ref.~\cite{ZavyalovEtAl2019} and in the authors' review paper~\cite{MartinVergaraEtAl2018}.
The scattering of kinks and antikinks in non-integrable equations is generally characterized by the existence of a critical initial speed $v_{cr}$ for its initial head-on speed $v$.
For $v>v_{cr}$ the solitons either pass through or bounce off each other reappearing after collisions with a phase shift in their positions.
But for $v<v_{cr}$ solitons form a breather-like state (referred to as an oscillon or a bion) that generally decays slowly by radiating energy in small-amplitude waves; however, there is a series of windows in initial speed where the solitons are able to escape to infinity.
{The cause of these windows is usually the resonance energy exchange between solitons kinetic energy and their internal vibrational (shape) modes};
however, there are counterexamples, nonlinear Klein--Gordon  equations without internal modes~\cite{DoreyEtAl2011}.
The observation of the quasi-fractal structure of the resonance windows is a clear indication that the solitary waves are not true solitons and that the equation is non-integrable.

The quasi-fractal structure of the resonance windows in the inelastic scattering of kinks and antikinks could be understood by recurring to a particle-wave analogy, the so-called resonant energy exchange theory developed by Campbell et al.~\cite{CampbellEtAl1983,Peyrard1983,CampbellEtAl1986}; these pioneering works applied this idea to the $\phi^4$ model, a modified sine-Gordon, and the double sine-Gordon equations, the last also studied by other authors~\cite{GaniKudryavtsev1999,GoodmanHaberman2005,GoodmanRahman2015,GaniEtAl2018, GaniEtAl2019}. In recent years other models have been considered; for example, a modified periodic version of the $\phi^4$ model~\cite{MohammadiDehghani2021}, the $\phi^6$ model~\cite{DoreyEtAl2011,GaniEtAl2014,MoradiEtAl2017}, the $\phi^8$ model~\cite{GaniEtAl2015}, the $\phi^{2\,m+4}$ models with $m=2, 3, 4, \ldots$~\cite{Christov2020, ChristovEtAlMay2019}, the sinh-deformed  $\phi^4$ model~\cite{BazeiaEtAl2018}, a weakly interacting $\phi^4$ model~\cite{AdamEtAl2020}, a biharmonic $\phi^4$ model~\cite{DeckerEtAl2020,TsoliasEtAl2020}, and a two-component $\phi^4$ model~\cite{AlonsoIzquierdo2020}, among others. Although the vast amount of evidence accumulated by the years on the validity of the resonant energy exchange theory, it remains as a phenomenological approach that requires rigorous support even for the $\phi^4$ model~\cite{Goodmanphi4chapter2019,Manton2021}. In our opinion, its application to  modifications of the sine-Gordon equation, like the graphene superlattice equation,  which can be studied by using perturbation techniques based on the inverse scattering technique, could help to clarify the limits of this theory; obviously, the first step in this direction is a comprehensive numerical analysis.

The analysis of the kink-antikink collisions in the GSLeq requires the use of an efficient method due to the large amount of numerical simulations needed. The authors have recently compared eight implicit finite difference methods with Pad\'e approximations in space for the sine-Gordon equation (sGeq), without~\cite{MartinVergaraEtAl2019} and with~\cite{MartinVergaraEtAl2020} Richardson extrapolation; these methods were inspired in the conservative methods developed by Strauss and V\'azquez~\cite{StraussVazquez1978} and Guo Ben--Yu et al.~\cite{GuoBenYuEtAl1986}. Our comparison among these Pad\'e methods indicates that the most efficient method for small global error in terms of accuracy and computational cost is a (4,0)-Pad\'e method based on the Guo Ben--Yu et al. method without Richardson extrapolation~\cite{MartinVergaraEtAl2020}. Hence this method will be used for the solution of the GLSeq under periodic boundary conditions.

The main goal of this paper is the numerical study of the kink-antikink collisions in the graphene superlattice equation. The contents of this paper are as follows. Section~\ref{sec:gsl} presents the graphene superlattice equation and shows the shape of their kinks and antikinks. Section~\ref{sec:schemes} reviews the numerical Pad\'e approximation scheme used in our simulations. Our  results are presented in Section~\ref{sec:numerical:results}; the initial condition for the kink-antikink collision is imbricated in order to cope with the periodic boundary conditions as shown in Subsection~\ref{sub:initialcondition}; the main results on the kink-antikink interaction are presented in Subsection~\ref{sub:interaction}; and the resonant energy exchange theory is exposed in Subsection~\ref{sub:resonant}. Finally, Section~\ref{sec:conclusions} is devoted to the conclusions and future research.

\section{Graphene superlattice equation}
\label{sec:gsl}

A strong electromagnetic (EM) field normally incident on the GSL induces the propagation of EM waves. The dimensionless component of the vector potential along the graphene superlattice axis, $\alpha$, solves the nonlinear Klein--Gordon equation given by~\cite{KryuchkovKukhar2013}
\begin{equation}
 \frac{\partial^2\alpha}{\partial t^2}
 - c^2\,\frac{\partial^2\alpha}{\partial x^2}
 + \frac{\omega_{pl}^2\,b^2\,\sin\alpha}
        {\sqrt{1+b^2\,(1-\cos\alpha)}}
 = 0,\label{martinvergara-KKeq}
\end{equation}
where $b$ is a geometrical parameter, and $\omega_{pl}$ is the plasma frequency. Hereafter, Eq.~\eqref{martinvergara-KKeq} is referred to as Graphene Superlattice Equation.

 {Under the condition $b^2/(1+b^2)\ll 1$, Eq.~\eqref{martinvergara-KKeq} reduces to the double sine-Gordon equation~\cite{KryuchkovChaotic2013}
\begin{equation}
 \frac{\partial^2\alpha}{\partial t^2}
 - c^2\,\frac{\partial^2\alpha}{\partial x^2}
 + \omega_{0}^2\,(\sin\alpha+\epsilon\,\sin 2\,\alpha)
 = 0,\label{martinvergara-dSGeq}
\end{equation}
where $\omega_{0}^2=\omega_{pl}^2\,b^2/\sqrt{1+b^2}$, and $\epsilon=b^2/(4\,(1+b^2))$. Equation~\eqref{martinvergara-dSGeq} is a conservative perturbation of the sGeq whose soliton scattering can be studied for small $\epsilon$ by perturbation methods based on the inverse scattering transform~\cite{KivsharMalomed1989}; for example, the production of breather-like solutions in the kink-antikink scattering for Eq.~\eqref{martinvergara-dSGeq} was analyzed by Malomed~\cite{Malomed1985}.}

A nondimensionalization of Eq.~\eqref{martinvergara-KKeq}, by the change of variables $t'={\omega_{pl}\,b\,t}$,  $x'={\omega_{pl}\,b\,x/c}$, and $u=\alpha$, yields
\begin{equation}
 \frac{\partial^2{u}}{\partial{t^2}}
  -\frac{\partial^2{u}}{\partial{x^2}}
  + \frac{dG(u)}{du} = 0,
 \qquad
 \frac{dG(u)}{du} = \frac{\sin u}{\sqrt{1+b^2\,(1-\cos u)}},
 \label{martinvergara-kkeq}
\end{equation}
where the primes have been dropped and
\begin{align}
   G(u) &= \frac{2\,(1-\cos u)}
        {1+\sqrt{1+b^2\,(1-\cos u)}}.
\end{align}
Note that the sGeq, i.e. $G(u) = 1 - \cos u$, is obtained for $b={0}$. A solitary wave solution of Eq.~\eqref{martinvergara-kkeq} with speed $v$ is obtained by Lorentz boosting a static solution $u(x)$, i.e. $u(x,t)=u((x-v\,t)/\sqrt{1-v^2})$, that solves
\begin{equation}
 \frac{d^2{u}}{dx^2} = \frac{dG(u)}{du},
 \qquad
 \frac{1}{2}\,\left(\frac{du}{dx}\right)^{2} = G(u).
  \label{martinvergara-kkeq:ode}
\end{equation}
The periodic potential $G(u)$ has infinite zeroes at $u^*_n = 2\,n\,\pi$, for $n\in\mathbb{Z}$; hence the static solitary waves connect two consecutive zeroes, being a $n$-kink solution when connecting the asymptotic value $u^*_n$, and $u^*_{n+1}$, and a $n$-antikink when connecting $u^*_{n+1}$, and $u^*_n$. In implicit form, the static $0$-kink, $u_{k,0}(x)$, is given by
\begin{equation}
 \int_{\pi}^{u_{k,0}} \frac{d\tilde{u}}{\sqrt{2\,G(\tilde{u})}} = \int_{x_0}^x d\tilde{x}, \qquad 0 < {u_{k,0}}  < 2\,\pi, \label{martinvergara-kkeq:static:solution}
\end{equation}
where $x_0$ is the center of the $0$-kink, i.e., the position of the maximum of its spatial derivative,
{$du_{k,0}/dx = \sqrt{2\,G(u_{k,0})}$; it is determined as $u_{k,0}(x_0)=\pi$, since this value nullifies the second-order spatial derivative, $d^2u_{k,0}(x_0)/dx^2 = dG(\pi)/du=0$. Moreover, the spatial derivative of the $0$-kink decays exponentially as $|x|\rightarrow\infty$, since its second spatial derivative decays linearly, $d^2u_{k,0}/dx^2 = u_{k,0} + \OL{u_{k,0}}^3$,  as $|x|\rightarrow\infty$, cf. Eq.~\eqref{martinvergara-kkeq}. Note also that} the $n$-kink can be written as $u_{k,n}(x)=2\,\pi\,n+u_{k,0}(x)$, and the $n$-antikink as $u_{ak,n}(x)=2\,\pi\,(n+1)-u_{k,0}(x)$.

For the GSLeq, Eq.~\eqref{martinvergara-kkeq:static:solution} for $u_{k,0}(x)$ reads as
\begin{equation}
 \int_{\pi}^{u_{k,0}}  \frac{\sqrt{1+\sqrt{1+b^2\,(1-\cos \tilde{u})}}}{2\,\sqrt{1 - \cos \tilde{u}}}\,d\tilde{u}
  = x - x_0,
 \label{martinvergara-kkeq:static:implicit}
\end{equation}
whose integration results in a cumbersome implicit expression written in terms of elliptic integrals. In practice, its numerical evaluation is used.

The application of a quadrature method to Eq.~\eqref{martinvergara-kkeq:static:implicit} has to take into account that the left-hand integrand is singular at $u_{k,0} =2\,\pi$, and $u_{k,0} =0$. In order to avoid such singularities, let us insert into Eq.~\eqref{martinvergara-kkeq:ode} the ansatz $u_{k,0}(x) = 4\,\arctan(\exp(w_{k,0}(x)))$, resulting in
\begin{equation}
 \frac{1}{\sqrt{2}}
 \int_{0}^{w_{k,0}}  \sqrt{1+\sqrt{1+2\,b^2\,\operatorname{sech}^2(\tilde{w})}}
 \,d\tilde{w}
  = x - x_0,
 \label{martinvergara-kkeq:static:newimplicit}
\end{equation}
where the identity
\[
 \cos(4\,\arctan(z)) = \frac{z^4 - 6\, z^2 + 1}{(z^2 + 1)^2},
\]
has been used.

\begin{figure}[t] \centering
\includegraphics*[width=\textwidth]{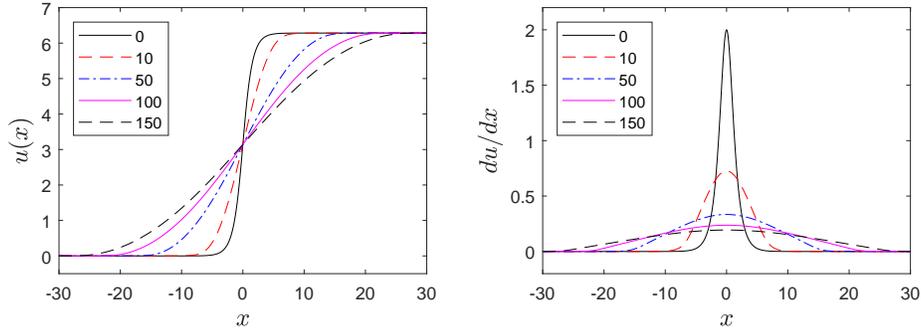}
\caption{Kink solution (left plot) and its spatial derivative (right plot) numerically calculated for $b=0$, $10$, $50$, $100$, and $150$, using $\Delta x=0.01$, and $x\in[-50, 50)$; the spatial derivative has been calculated by using the fast Fourier transform.} \label{martinvergara-figure1}
\end{figure}

Figure~\ref{martinvergara-figure1} shows the kink solution for $b=0$, $10$, $50$, $100$, and $150$ (notice that in Ref.~\cite{MartinVergaraEtAl2018} it  has been plotted for $b\le 1$). It has been numerically calculated by applying a Gauss--Konrod quadrature formula for the left-hand side of Eq.~\eqref{martinvergara-kkeq:static:newimplicit}, and by using an inverse interpolation technique based on cubic splines for solving the nonlinear equation. The plots in Fig.~\ref{martinvergara-figure1} show that the solution widens as the parameter $b$ grows, starting with the sharpest one for $b=0$, i.e., the kink of the sGeq.


\section{Numerical scheme}
\label{sec:schemes}

Let us consider the numerical solution of the GSLeq by means of the most effective method for the sGeq among those Pad\'e methods compared by the authors in Ref.~\cite{MartinVergaraEtAl2020}. This method is inspired in the energy conservation scheme by Guo Ben-Yu et al.~\cite{GuoBenYuEtAl1986} for the nonlinear Klein--Gordon equation; it uses the same second-order, leap-frog stencil in time, and the same Strauss and V\'azquez~\cite{StraussVazquez1978} treatment for the nonlinear term, but with a fourth-order, central difference formula in space.

The numerical scheme can be written as
\begin{align}
    \frac{U_m^{n+1}-2\,U_m^n+U_m^{n-1}}{\Delta t^2}
  - \mathcal{A}(\mbox{E})\,\frac{({U_m^{n+1}+U_m^{n-1}})}{2}
  +  
    H(U_m^{n+1})= 0,
 \label{eq:method}
\end{align}
with
\begin{equation}
 {\mathcal{A}(\mbox{E})}
 = \frac{ -\mbox{E}^{-2}+16\,\mbox{E}^{-1}-30
          +16\,\mbox{E}^{1}-\mbox{E}^{2}
     }{12\, \Delta x^2},
 \label{eq:fdm4}
\end{equation}
and
\[
 H(U_m^{n+1}) \equiv \frac{G(U_m^{n+1})-G(U_m^{n-1})}{U_m^{n+1}-U_m^{n-1}},
\]
where $U_m^n \approx u(x_m, t^n)=u_m^n$, $x_m=m\,\Delta x$, for $m\in\mathbb{Z}$, $\Delta x$ is the grid size, $t^n=n\,\Delta t$, for $n\in\mathbb{N}$, $\Delta t$ is the time step, $\mathcal{A}(\mbox{E})\,u_m^n$ is a fourth-order discretization of the spatial derivative, and $\mbox{E}$ is the shift operator defined as $\mbox{E}\,U_m^n = U_{m+1}^n$. We use periodic boundary conditions in the finite interval $x \in (-L, L]$, with $x_m=-L+m\,\Delta x$, $m=1, 2, \ldots, M$, and $\Delta x = 2\,L/M$ (note that $u(x_{0}, t) \equiv u(x_{M}, t)$), and a finite time interval $t\in[0, T]$, with $t^n=n\,\Delta t$, $n=0, 1, \ldots, N$, and $\Delta t=T/N$.

Method~\eqref{eq:method} is implicit, since the calculation of $U_m^{n+1}$ from $U_m^n$ and $U_m^{n-1}$ requires the solution of a nonlinear equation. Let us use Newton's iterative method given by
\begin{align}
  & {U_m^{(k+1)}-2\,U_m^n +U_m^{n-1}}
  - \frac{\Delta t^2}{2}\, {\mathcal{A}(\mbox{E})}\,({U_m^{(k+1)}+U_m^{n-1}})
  \nonumber \\ & \qquad  +
  {\Delta t^2}\,
   \left(
    H(U_m^{(k)}) +
    H_u(U_m^{(k)})\,(U_m^{(k+1)}-U_m^{(k)})
   \right) = 0,
 \label{eq:method:NM}
\end{align}
with
\[
 H_u(U_m^{(k)}) \equiv
   \frac{G_u(U_m^{(k)})\,(U_m^{(k)}-U_m^{n-1})
         -(G(U_m^{(k)})-G(U_m^{n-1}))}
        {\left(U_m^{(k)}-U_m^{n-1}\right)^2} .
\]

Method~\eqref{eq:method} is linearly, unconditionally stable, highly accurate and has good energy conservation properties even for $\Delta t=\Delta x$, as shown by the authors in Ref.~\cite{MartinVergaraEtAl2020}.


\section{Presentation of results}
\label{sec:numerical:results}

Let us summarize the main results for the behaviour of the GSLeq, obtained after a large set of simulations. In Subsection~\ref{sub:initialcondition} the initial condition for the kink-antikink solution is presented; Subsection~\ref{sub:interaction} shows the results for the kink-antikink interaction and in Subsection~\ref{sub:resonant} the resonant energy exchange theory is studied.

\subsection{Kink-antikink initial condition}
\label{sub:initialcondition}

The exact solution of Eq.~\eqref{martinvergara-kkeq} for the interaction between a kink and an antikink is not known. A good approximation can be obtained by using the so-called \emph{sum ansatz}~\cite{ChristovEtAl2019} given by
\begin{equation}
 \tilde{u}_{kak}(x,t) =
   u_{k}(\frac{x+x_0-v\,t}{\sqrt{1-v^2}})+
   u_{ak}(\frac{x-x_0+v\,t}{\sqrt{1-v^2}}),
 \label{martinvergara-kkeq:kinkantikinknoimbricated}
\end{equation}
where the kink is located at $x_0$ and the antikink at $-x_0$.
In our numerical simulations, periodic boundary conditions are used; but Eq.~\eqref{martinvergara-kkeq:kinkantikinknoimbricated} is not a periodic solution of Eq.~\eqref{martinvergara-kkeq}, having non-continuous derivatives at the boundaries. In order to solve this problem an imbricated soliton series can be used~\cite{Boyd1986,Boyd1989}. The imbrication for the kink-antikink solution yields
\begin{equation}
 u_{kak}(x,t) = \sum_{j=-p}^{p}
   \tilde{u}_{kak}(x+2\,L\,j,t), \qquad x\in [-L,L),
 \label{martinvergara-kkeq:kinkantikink}
\end{equation}
for $p\rightarrow\infty$; in fact, when $x_0=L/2$, and $L$ is larger than the width of the kink, our experience indicates that $p=1$ results in an accurate periodic solution since the spatial derivatives of the kink and the antikink are exponentially decaying, {cf. the discussion after Eq.~\eqref{martinvergara-kkeq:static:solution}}. 

The solution $u_{kak}(x,t)$ approximately satisfies Eq.~\eqref{martinvergara-kkeq} for $x_0$ larger than the width of the kink, and $t$ small enough such that the kink and the antikink are well-separated. In Ref.~\cite{ChristovEtAl2019} the accuracy has been tested by calculating the residual of Eq.~\eqref{martinvergara-kkeq}, i.e.,
\[
 \text{Res}(t) = \max_{x\in[L,L)} \left|\frac{\partial^2{}}{\partial{t^2}}u_{kak}(x,t)
  -\frac{\partial^2{}}{\partial{x^2}}u_{kak}(x,t)
  + \frac{dG}{du}(u_{kak}(x,t))
 \right|.
\]
In the numerical simulations presented in this paper, Eq.~\eqref{martinvergara-kkeq:kinkantikink} is only used for $t\le{0}$; since the kink sharpens as $v$ grows, the largest value of $\text{Res}(t)$ is obtained for $v=0$; hence, the numerical accuracy of $u_{kak}(x,t)$ can be tested at $t=0$ by calculating the time-independent residuals of Eq.~\eqref{martinvergara-kkeq:ode}, i.e.,
\begin{equation}
 \text{Res}_1 = \max_{x\in(-L,L]}  \left|\frac{d u_{kak}(x,0)}{dx} -
         \sqrt{2\,G(u_{kak}(x,0))}
       \right|,
 \label{martinvergara-kkeq:residual:first-order}
\end{equation}
and
\begin{equation}
 \text{Res}_2 = \max_{x\in(-L,L]}  \left| \frac{d^2{u_{kak}(x,0)}}{dx^2} -
               \frac{dG(u_{kak}(x,0))}{du}
       \right|.e
 \label{martinvergara-kkeq:residual:second-order}
\end{equation}

Table~\ref{tab:residuals} shows the numerical evaluation of residuals~\eqref{martinvergara-kkeq:residual:second-order} and~\eqref{martinvergara-kkeq:residual:first-order} as a function of $b=0, 1, \ldots, 200$, for $x_0=25$ and $x_0=50$, for $x\in(-L, L]$, with $L=2\,x_0$. For $x_0=25$, the residual $r_1$ is approximately constant of the order of $10^{-8}$ for $b < 25$, but monotonically grows for $b\ge 25$, reaching unacceptable large errors for $b>75$; and the residual $r_2$ has a similar behaviour. For $x_0=50$, the residual $r_1$ is equal to $1.05 \times 10^{-8}$ for all $b$, but $r_2$ first grows for $b\le 10$, and then decreases, but always remains of the order of $10^{-7}$. Hence, in numerical simulations we recommend the use of $x_0=25$ for $b<25$, and $x_0=50$ for $25 \le b \le 150$; for larger values $b$, an even larger $x_0$ is required for a small enough residual.

\begin{table}
\centering
\caption{Residuals as function of $b$ calculated using Eqs.~\eqref{martinvergara-kkeq:residual:second-order} and~\eqref{martinvergara-kkeq:residual:first-order} for the kink-antikink solution with $x_0=25$ and $50$, when using $\Delta x$   with $L=4\,x_0$ .}
\begin{tabular}{c|ll|ll}
 $b$ & \multicolumn{2}{c|}{$x_0 = 25$}
 & \multicolumn{2}{c}{$x_0 = 50$}
\\ \hline
 & \multicolumn{1}{c}{$\text{Res}_1$} & \multicolumn{1}{c|}{$\text{Res}_2$} & \multicolumn{1}{c}{$\text{Res}_1$} & \multicolumn{1}{c}{$\text{Res}_2$}
\\ 
$0$ &  $1.05 \times 10^{-8}$ & $2.46 \times 10^{-10}$  & $1.05 \times 10^{-8}$ & $3.48 \times 10^{-10}$
\\ 
$1$ &  $1.05 \times 10^{-8}$ & $4.98 \times 10^{-8}$   & $1.05 \times 10^{-8}$ & $4.96 \times 10^{-8}$
\\ 
$10$ & $1.03 \times 10^{-8}$ & $3.33 \times 10^{-7}$   & $1.05 \times 10^{-8}$ & $3.33 \times 10^{-7}$
\\ 
$25$ & $4.94 \times 10^{-8}$ & $2.72 \times 10^{-7}$   &  $1.05 \times 10^{-8}$ & $2.72 \times 10^{-7}$
\\ 
$50$ & $2.39 \times 10^{-6}$ & $1.94 \times 10^{-7}$   & $1.05 \times 10^{-8}$ & $1.94 \times 10^{-7}$
\\ 
$75$ & $5.34 \times 10^{-5}$ & $1.27 \times 10^{-6}$   & $1.05 \times 10^{-8}$ & $1.44 \times 10^{-7}$
\\ 
$100$ & $7.72 \times 10^{-4}$ & $2.37\times 10^{-5}$   & $1.05 \times 10^{-8}$ & $1.16 \times 10^{-7}$
\\ 
$125$ & $ 7.93\times 10^{-3}$ & $2.49 \times 10^{-2}$  & $1.05 \times 10^{-8}$ & $9.81 \times 10^{-8}$
\\ 
$150$ & $5.63 \times 10^{-2}$ & $4.87 \times 10^{+0}$  & $1.05 \times 10^{-8}$ & $8.44 \times 10^{-8}$
\\ 
$175$ & $4.04 \times 10^{-1}$ & $66.3 \times 10^{+0}$  & $1.05 \times 10^{-8}$ & $7.04 \times 10^{-8}$
\\ 
$200$ & $2.15 \times 10^{+0}$ & $399 \times 10^{+0}$  & $1.05 \times 10^{-8}$ & $6.60 \times 10^{-8}$
\\ \hline
\end{tabular}
\label{tab:residuals}
\end{table}

\subsection{Kink-antikink interaction}
\label{sub:interaction}

The GSLeq is not integrable, since it has not the Painlev\'e property~\cite{ClarksonEtAl1986}; hence its kinks and antikinks are not solitons, in strict sense, but they are solitary waves showing radiative tails after their mutual interaction, as numerically confirmed in Ref.~\cite{ZavyalovEtAl2019}. Similarly to other non-integrable, nonlinear Klein--Gordon equations~\cite{CampbellEtAl1983}, when using Eq.~\eqref{martinvergara-kkeq:kinkantikinknoimbricated} as initial conditions with a given initial velocity, the long-time behaviour of the solution depends on whether this initial velocity is smaller or bigger than some critical velocity $v_{cr}$ that is a function of the parameter $b$.

\begin{figure}[t] \centering
\includegraphics[width=\textwidth]{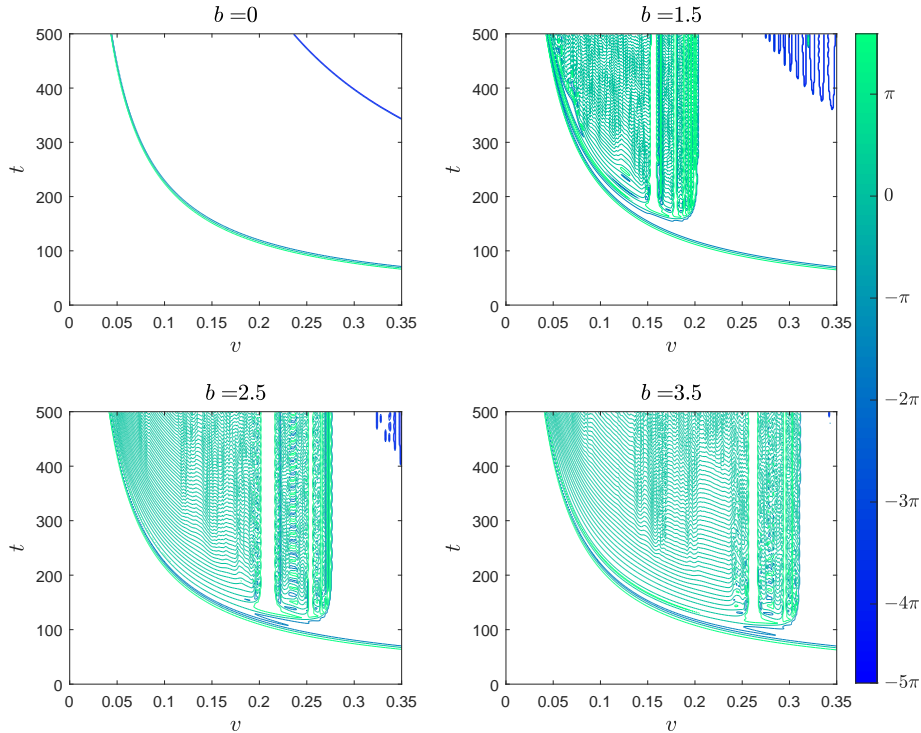}
\caption{Contour plots of the kink-antikink collision  $u(0,t;v)$ in the plane $(v,t)$, with contour levels in $\{-15, -10, -5, 0, 5\}$, and $v\in[0,0.35]$ in velocity steps of $\Delta v=10^{-3}$, for $b = 0$, 1.5, 2.5, and 3.5, with $x_0=25$, $\Delta t = \Delta x = 0.01$, $L=100$, and $T=500$. Note that the multiples of $\pi$ has not been used as contour levels in order to reduce the noise in the plot. }
\label{fig:F2}
\end{figure}

Let us write as $u(x,t;v)$ the solution of the GSLeq when the initial condition is $\tilde{u}_{kak}(x,0)$ with speed $v$. Figure~\ref{fig:F2} shows contour plots of the solution $u(0,t;v)$ in the plane $(v,t)$, with $v\in[0,0.35]$ in velocity steps of $\Delta v=10^{-3}$, contour levels at $\{-15, -10, -5, 0, 5\}$, for $b = 0$, 1.5, 2.5, and 3.5, with $x_0=25$, $\Delta t = \Delta x = 0.01$, $L=100$, and $T=500$.
For $b=0$, the sGeq under periodic boundary conditions, the top left plot shows two groups of contour curves for $u(0,t;v)$ with levels equal to $\{-5, 0, 5\}$ for the first kink-antikink collision (green color) and equal to $\{-15, -10\}$ for the second one (blue color). An analytical expression for these contour level curves can be easily obtained by generalizing Eq.~(2.2) in Ref.~\cite{CampbellEtAl1983}; the  time for the $q$-th collision, such that $u(0,t_q;v)=-4\,\pi\,(q-1)$, is given by
\begin{equation}
  t_q(v) = \frac{x_0+L\,(q-1)}{v}
     + \frac{(1+4\,(q-1))}{v}\,(1-v^2)^{1/2}\,\ln(v),
  \label{contour:curves}
\end{equation}
for $x_0<L/2$.
For $b>0$, the contour curves associated with the first collision are well described by Eq.~\eqref{contour:curves} with $q=1$. However, for the second collision, the contours can be approximated by Eq.~\eqref{contour:curves} with $q=2$ only for $b<1$; let us highlight that the critical velocity for $b=1$ is $v_{cr}=0.138$. For $b>1$, the critical velocity approaches the position of the contours for the second collision, shifting their positions to higher values. In order to illustrate such a behaviour, the contour plots for $b=1.5$, 2.5, and 3.5 are shown in the top right, bottom left, and bottom right plots in Fig.~\ref{fig:F2}, respectively.

\begin{figure}[t]  \centering
\includegraphics[width=\textwidth ]{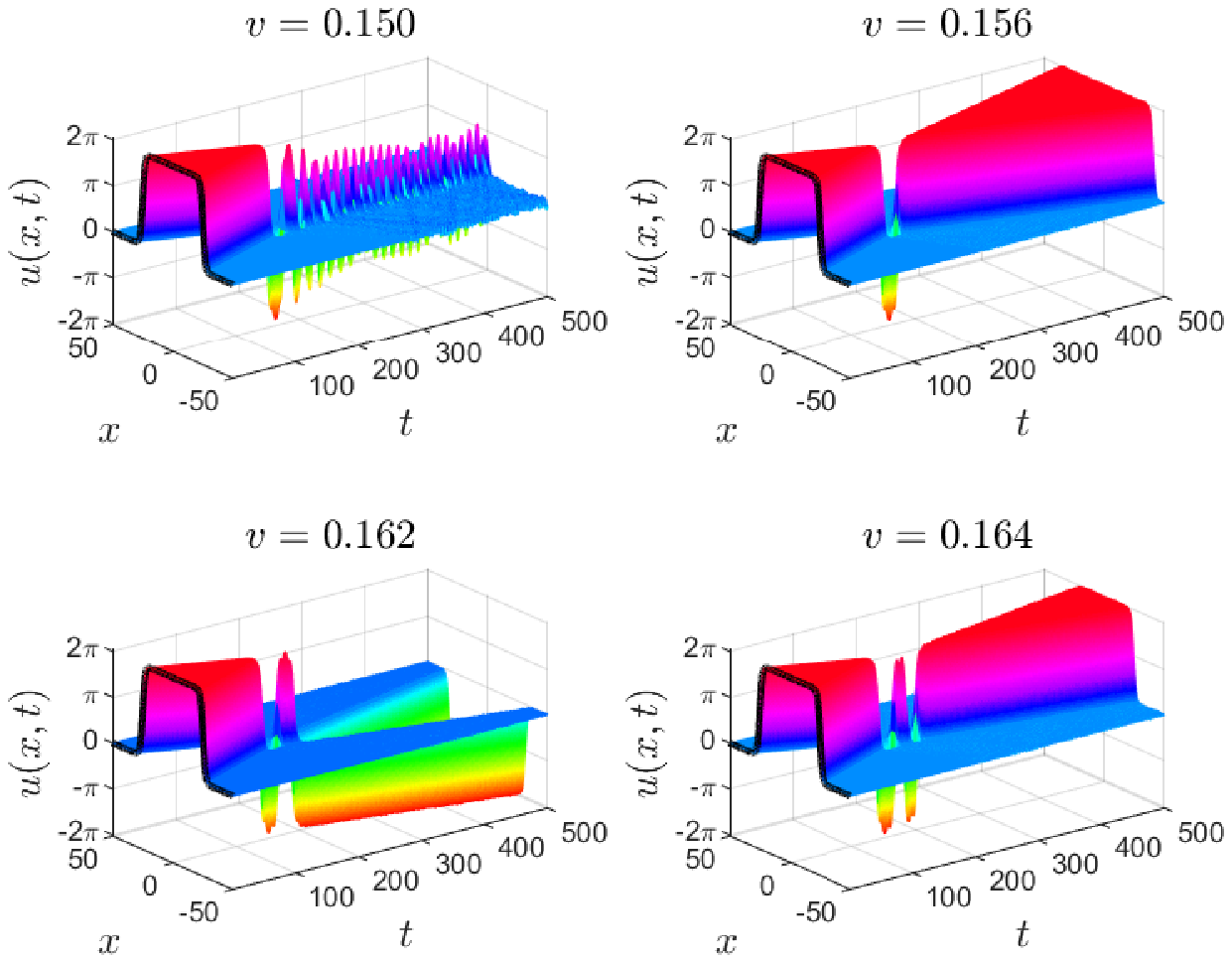}
\caption{Mesh plots of the kink-antikink collision $u(x,t;v)$ for $v=0.150$ (top left), $0.156$ (top right), $0.162$ (bottom left plot), and $0.164$ (bottom right),
with $b=1.5$, $x_0=25$, $\Delta t = \Delta x = 0.01$, $L=100$, and $T=500$.}
\label{fig:F3}
\end{figure}

Figure~\ref{fig:F3} shows a mesh plot of the kink-antikink collision $u(x,t;v)$ for $b=1.5$ with $v=0.150$ (top left), $0.156$ (top right), $0.162$ (bottom left), and $0.164$ (bottom right); note that all these initial speeds are smaller than the corresponding critical velocity $v_{cr}=0.20405$. The kink and the antikink can be trapped in an oscillatory bound state, a breather-like state referred to as an oscillon or a bion, with slowly diminishing amplitude as time increases, as shown in Fig.~\ref{fig:F3} (top left).
This general behaviour is not observed for all the initial speeds below the critical velocity, since there are a series of resonance windows in initial velocity, where after several bounces the kink and the antikink escape to infinity; Fig.~\ref{fig:F3} (top right) shows an example of a two-bounce scattering, Fig.~\ref{fig:F3} (bottom left) of a three-bounce one, and Fig.~\ref{fig:F3} (bottom right) of four-bounce one. Note that the scattering within the resonance windows is inelastic, since the final velocity is smaller than the initial velocity.
Further plots obtained after an extensive set of results show that the initial kink-antikink pair after an odd number of bounces escapes as an antikink--kink pair (they pass through each other), but after an even number of bounces escapes as a kink-antikink pair (they bounce off each other). Note that the multi-bounce scattering is a general prediction of the theory of chaotic scattering of solitons. This theory predicts  an infinite numerable set of two-bounce windows of increasing initial speed $v<v_{cr}$ where the kink and antikink are temporally trapped in a resonance state before finally escaping from each other's influence, and also that the output velocity $v_{out}$ of the kinks is always smaller than the input velocity $v$ inside the two-bounce windows~\cite{GoodmanHaberman2007,Goodman2008}.

\begin{figure}[t]
\begin{center}
\includegraphics[width=\textwidth]{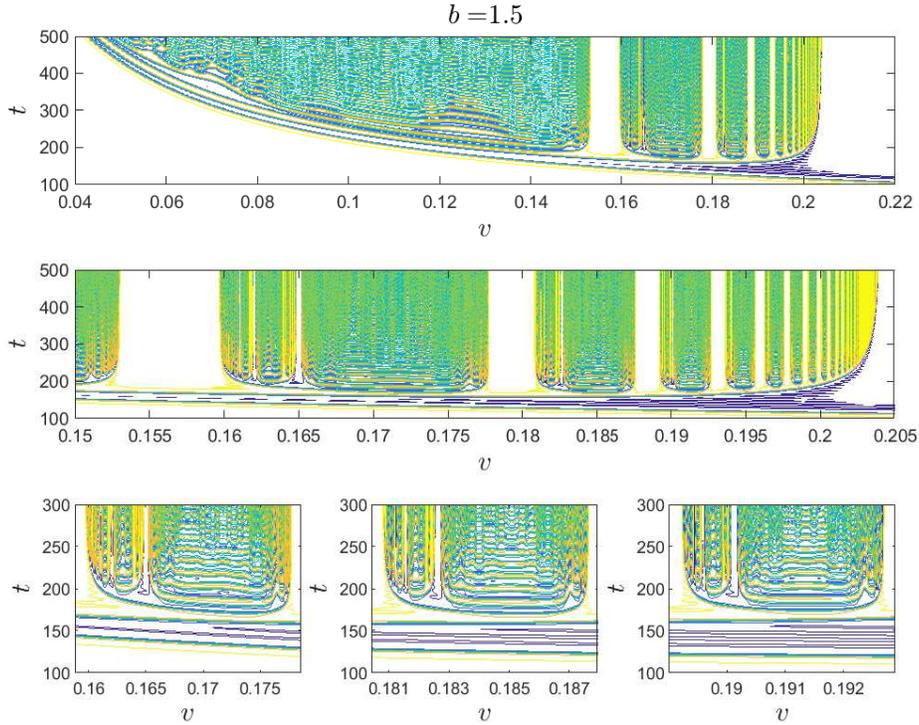}
\end{center}
\caption{The top plot shows the kink-antikink collision  $u(0,t;v)$ for $b=1.5$ contour plotted in the plane $(v,t)$ for $v\in [0.04,0.22]$, with $\Delta v=10^{-3}$ when $v\le 0.15$ and $\Delta v=10^{-4}$ otherwise, and $t\in[100,500]$, by using simulations $\Delta t = \Delta x = 0.01$. The middle plot shows a zoom of the contour plot $u(0,t;v)$ for $v\in [0.150,0.205]$, with $\Delta v=10^{-5}$, and $t\in[100,500]$, using simulations with $\Delta x = 0.1$ and $\Delta t = 0.01$. And the bottom plots show zooms of $u(0,t;v)$ with $t\in[100,300]$, for $v\in [0.15880, 0.17858]$ (bottom left), $v\in [0.18044, 0.18797]$ (bottom center), and $v\in [0.18901, 0.19289]$ (bottom right), using simulations with $\Delta x = 0.1$ and $\Delta t = 0.01$. In all the simulations $x_0=25$, $L=100$, and $T=500$. }
\label{fig:F4}
\end{figure}

Figure~\ref{fig:F4} (top) shows the contour plot of the kink-antikink collision $u(0,t;v)$ for $b = 1.5$ in the plane $(v,t)$ with $v\in [0.04,0.22]$, and $t\in[100,500]$; note that in the horizontal axis $\Delta v=10^{-3}$ was used for $v\le 0.15$ and $\Delta v=10^{-4}$ otherwise, and that the simulations used $\Delta t = \Delta x = 0.01$, $x_0=25$, $L=100$, and $T=500$. The contour plot in Fig.~\ref{fig:F4} (top) shows a series of white bands centered around $v_i=0.156$, 0.179, 0.188, 0.193, \ldots, corresponding to the $i$-th two-bounce resonance window; note that only the first and second ones are noticeable in the top right plot in Fig.~\ref{fig:F2}. The width of the $i$-th two-bounce window, $W_i=0.007$, 0.003, 0.002, 0.001, \ldots, decreases to zero as their centers approach $v_{cr}$; hence, the critical velocity as an accumulation point for these resonance windows.

The kink-antikink collisions in Fig.~\ref{fig:F4} (top) apparently shows a self-similar behaviour in the regions between the two-bounce resonance windows; note that it does not occur before the first one, i.e., for $v<0.153$. In order to highlight this self-similarity, Fig.~\ref{fig:F4} (middle) shows a zoom of the contour plot of $u(0,t;v)$ for $v\in [0.150,0.205]$, and $t\in[100,500]$, and Fig.~\ref{fig:F4} (bottom) shows three zooms for $v\in [0.15880, 0.17858]$ (bottom left), $v\in [0.18044, 0.18797]$ (bottom center), and $v\in [0.18901, 0.19289]$ (bottom right), with $t\in[100,300]$, and $\Delta v=10^{-5}$. The comparison of the region between the first and second two-bounce windows, Fig.~\ref{fig:F4} (bottom left), with that between the second and third windows, Fig.~\ref{fig:F4} (bottom center), or with that between the third and fourth windows, Fig.~\ref{fig:F4} (bottom right), clearly confirms the self-similarity; successive zoomed contour plots, omitted for the sake of brevity, also present this self-similarity, although there is a lost of fine details as the width of the region between resonance windows diminishes. The three plots in Fig.~\ref{fig:F4} (bottom) show small multi-bounce windows with more than two bounces.

\begin{figure}[t]
\begin{center}
\includegraphics[width=\textwidth]{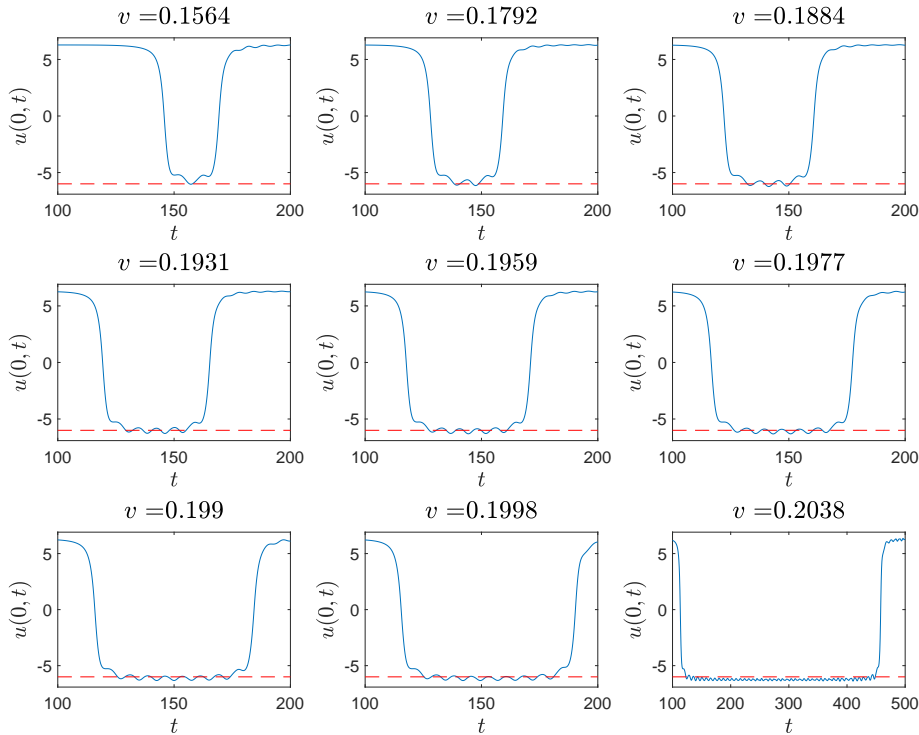}
\end{center}
\caption{The plot of $u(0,t;v)$ for several values of $v$ in the center of the first eight two-bounce windows and the last one marked with a circle in Fig.~\ref{fig:F7}~(top), for $b=1.5$, $x_0=25$, $\Delta t = \Delta x = 0.01$, $L=100$, and $T=500$. The red dashed line marks $u(0,t;v)=-6$.}
\label{fig:F5}
\end{figure}

In every two-bounce window, the function  $u(0,t;v)$ shows an oscillatory waveform with an integer number of oscillation peaks; it is determined by counting the number of local maxima of $|u(0,t;v)|$ for $t$ such that $|u(0,t;v)|>6$ inside the two-bounce window. The number of oscillations peaks is referred to as the index of the window in Ref.~\cite{CampbellEtAl1983}, where the $\phi^4$ model was studied; here on the two-bounce window with index $i$ is referred to as an $(i)$-window, for $i\ge 1$.
Figure~\ref{fig:F5}, from the top left plot to the bottom right plot, shows $u(0,t;v)$ with $t\in[100,200]$ for several two-bounce windows with $v=0.1564$, $0.1792$, \ldots, $0,1998$, and $0.2038$ corresponding to $(1)$-, $(2)$-, \ldots, $(8)$-, and $(45)$-windows, respectively.
The number of oscillation peaks between the bounces in the plots in Fig.~\ref{fig:F5} can also be observed in the bottom plots of Fig.~\ref{fig:F4}. The oscillation in the $(1)$-window corresponds to the double line around $t\approx 150$ for $v\approx 0.16$ in Fig.~\ref{fig:F4} bottom left plot; the two oscillations in the $(2)$-window to the two double lines around $t\approx 140$--$150$ for $v\approx 0.180$ in Fig.~\ref{fig:F4} bottom middle plot and for $v\approx 0.181$ in Fig.~\ref{fig:F4} bottom center plot; the three oscillations in the $(3)$-window to the three double lines around $t\approx 130$--$150$ for $v\approx 0.188$ in Fig.~\ref{fig:F4} bottom center plot and for $v\approx 0.189$ in Fig.~\ref{fig:F4} bottom right plot; and successively.

\begin{figure}[t]
\begin{center}
\includegraphics[width=\textwidth]{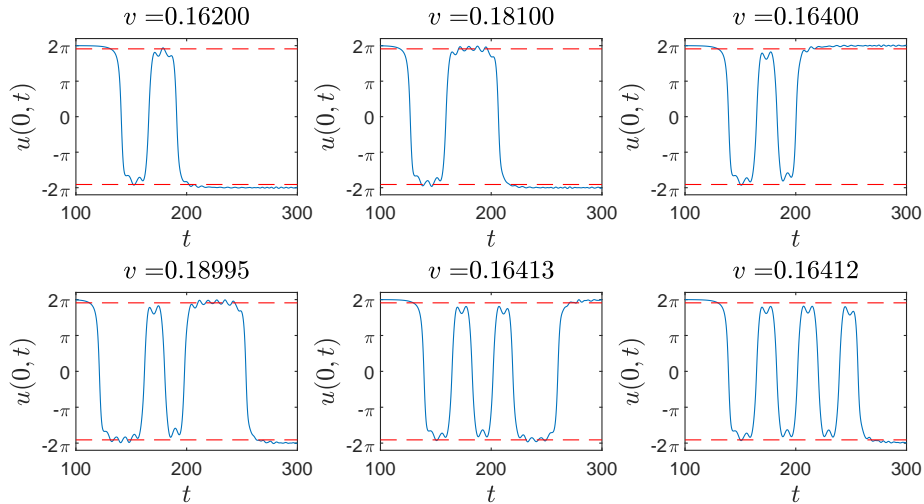}
\end{center}
\caption{Plots of $u(0,t;v)$ for  $(1_R,1)$-, $(2_R,4)$-, $(1_R,0,0)$-, $(3_R,0,0,5)$-, $(1_R,0,0,0,3)$-,  and $(1_R,0,0,0,0,0)$-windows,  with $\Delta x = 0.1$, $\Delta t = 0.01$, $x_0=25$, $L=100$, and $T=500$. See the main text for an explanation of the multi-index notations used. The red dashed lines mark $|u(0,t;v)|=6$.}
\label{fig:F6}
\end{figure}

Figure~\ref{fig:F6} shows plots of $u(0,t;v)$ for the central velocities of several multi-bounce windows for $t\in[100,300]$, concretely, for $v=0.16200$, $0.18100$, $0.16400$, $0.18995$, $0.16413$, and $0.16412$, from top left to bottom right. Notice that the top left and the top right plots in  Fig.~\ref{fig:F6}  correspond to the kink-antikink collision in Fig.~\ref{fig:F3} bottom left and bottom right mesh plots, respectively.
The top left plot in Fig.~\ref{fig:F6} shows a three-bounce window with one oscillation peak between the first and second bounces, and another one between the second and third bounces. The top center plot in Fig.~\ref{fig:F6} shows a three-bounce window with two oscillations peaks between the first and second bounces, and four ones between the second and third bounces. In some multi-bounce windows we observe the absence of oscillations peaks with $|u(0,t;v)|>6$ between some bounces; they show two peaks with $|u(0,t;v)|<6$ that do not correspond to the notion of index of the window introduced in Ref.~\cite{CampbellEtAl1983}; here on these features are referred to as non-oscillation peaks with an equivalent index of 0. Up to this authors' knowledge the existence of solutions with non-oscillation peaks between bounces have not been reported in the literature for other nonlinear Klein--Gordon equations.
The top right plot in Fig.~\ref{fig:F6} shows an example of non-oscillation peaks for a four-bounce window with one oscillation peak between the first and second bounces, a non-oscillation peak between the second and third bounces, and another one between third and fourth bounces.
The bottom plots in Fig.~\ref{fig:F6} show further examples of multi-bounce windows with two (left plot), three (center plot) and five (right plot) non-oscillation peaks.

In order to describe the self-similar structure of the multi-bounce windows, let us introduce a multi-index notation  that generalizes the index  introduced in Ref.~\cite{CampbellEtAl1983}. Let us anticipate that, for the  GSLeq, there can be  $p$-bounce windows around both sides of the $2$-bounce windows. Hence, the
$(i_{1\,L},i_2,\ldots,i_{p-1})$- and $(i_{1\,R},i_2,\ldots,i_{p-1})$-windows,
with $i_q\in \mathbb{N}$, are $p$-bounce windows with $i_q$ oscillations in each $q$-th bounce, which are located at the left and right sides, respectively, of the $(i_1)$-window; in order to shorten the multi-index notation, the non-oscillation peaks are represented by $i_q=0$.
Using this notation, Fig.~\ref{fig:F6} top left, center, and right plots correspond to $(1_R,1)$-, $(2_R,4)$-, and $(1_R,0,0)$-windows, resp., and bottom left, center, and right plots correspond to $(3_R,0,0,5)$-, $(1_R,0,0,0,3)$-, and $(1_R,0,0,0,0,0)$-windows, respectively.
Each $(i_{1\,L|R},i_2,\ldots,i_{p-1})$-window is characterized by its central velocity, denoted as $v_{(i_{1\,L|R},i_2,\ldots,i_{p-1})}$, its left boundary velocity, indicated as $v^{L}_{(i_{1\,L|R},i_2,\ldots,i_{p-1})}$, and its right boundary velocity, as $v^{R}_{(i_{1\,L|R},i_2,\ldots,i_{p-1})}$.

\begin{figure}[!t]
\begin{center}
\includegraphics[width=\textwidth]{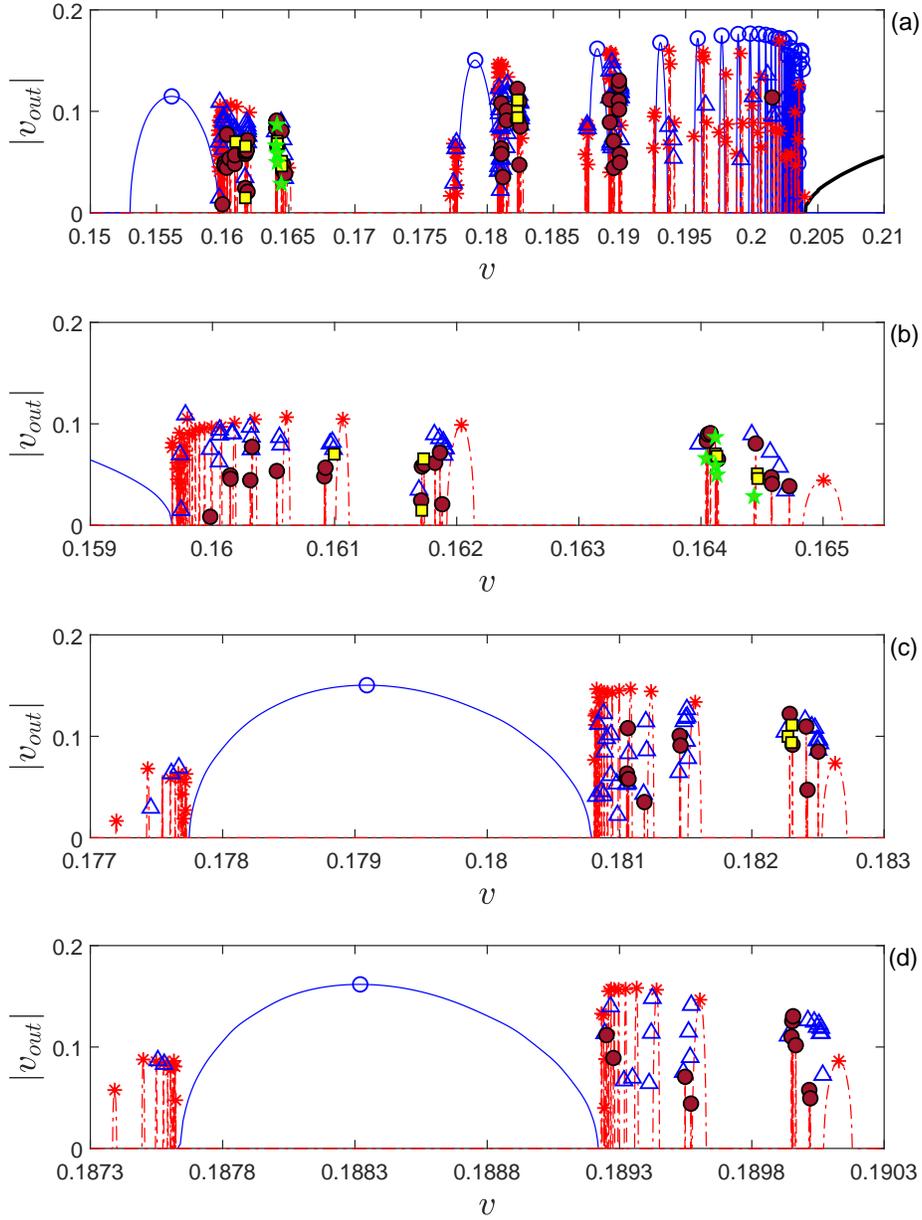}
\end{center}
\caption{The absolute value of the output velocity $|v_{out}|$ of the kink after the kink-antikink interaction as a function of its initial velocity $v$ for $b=1.5$ (a), and zooms in the intervals $v\in[0.155,0.166]$ (b), $v\in[0.177,0.183]$ (c), and $v\in[0.1873,0.1903]$ (d). See the main text for a detailed explanation of this figure. All the simulations use $\Delta x = 0.1$, $\Delta t = 0.01$, $x_0=25$, $L=100$, and $T=500$.  }
\label{fig:F7}
\end{figure}

Figure~\ref{fig:F7}~(a) shows the absolute value of the output velocity  $|v_{out}|$ of the kink after the kink-antikink collision as a function of its initial velocity $v\in[0.15,0.21]$ for $b=1.5$. In the horizontal axis for $v\in[0.15000,0.21000]$, a $\Delta v = 10^{-5}$ was generally used; however, in order to obtain a most detailed plot, a $\Delta v = 10^{-6}$ was used for  $v\in[0.159660,0.165200]$, $[0.176000,0.177800]$, $[0.180800,0.182800]$, $[0.187000,0.187630]$, and $[0.189200,0.190200]$. Note that $v_{out}>0$  ($v_{out}<0$) for the even-bounce (odd-bounce) windows, and that the 2-, 3-, 4-, 5-, 6-, and 7- windows are marked by circle, asterisk, triangle, filled-circle, square, and star symbols.
Let us emphasize that $v_{out}$ was numerically calculated by means of a linear interpolation in time of the position of the kink after the first kink-antikink collision, as seen in the plots in Fig.~\ref{fig:F3}. The interpolation interval in time was selected depending on the number of bounces of the solution;
for two-bounce solutions the interval $t\in[300, 350]$ was used for $v\in[0.15000,0.15965]$, $t\in[500, 700]$ was used for $v\in[0.159660,0.159699]$, $t\in[390,440]$ for $v\in[0.15970,0.15980]$, $t\in[300,390]$ for $v\in[0.159801, 0.203190]$, $t\in[480,580]$ for $v\in[0.20320,0.20378]$, and $t\in[500,700]$ for $v\in[0.20379,0.21000]$; for three-bounce solutions the interval $t\in[300, 350]$ was used for $v\in[0.15000,0.15965]$, $t\in[860, 940]$ was used for $v\in[0.159660,0.159699]$, $t\in[390,440]$ for $v\in[0.15970,0.15980]$, $t\in[300,390]$ for $v\in[0.159801, 0.203190]$, $t\in[480,580]$ for $v\in[0.20320,0.20378]$, and $t\in[500,700]$ for $v\in[0.20379,0.21000]$; and for solutions with more than three bounces the interval in time has been manually selected for an accurate linear interpolation after a visual inspection (the specific intervals are omitted here for the sake of brevity).

%
The continuous blue curves marked with a blue circle symbol in the plots in Fig.~\ref{fig:F7} are $(i_1)$-windows with $i_1\ge 1$, corresponding to the white bands shown in Fig.~\ref{fig:F4} (middle); the  solution inside such  windows is illustrated in the mesh plot in Fig.~\ref{fig:F3} (top right) for $v_{(1)}=0.156$, the center of the $(1)$-window, and also in the plots in Fig.~\ref{fig:F5}, corresponding to the $(1)$-, $(2)$-, \ldots, $(8)$-, and $(45)$-windows.
The dot-dashed red curves marked with a red asterisk symbol in the plots in Fig.~\ref{fig:F7} are $(i_{1 L|R},i_2)$-windows with $i_1\ge 1$ and $i_2\ge 0$; their typical evolution is shown in Fig.~\ref{fig:F3} (bottom left) for $v_{(1_R,1)}=0.162$, cf. the top left plot in Fig.~\ref{fig:F6}.
The blue triangle symbols in Fig.~\ref{fig:F7} correspond to the $(i_{1 L|R},i_2,i_3)$-windows with $i_1\ge 1$ and $i_2,i_3\ge 0$; their kink-antikink collision is illustrated in the mesh plot in Fig.~\ref{fig:F3} (bottom right) for $v_{(1_R,0,0)}=0.164$, cf. the top right plot in Fig.~\ref{fig:F6}.
In addition, the plots in Fig.~\ref{fig:F7} show brown filled-circle symbols corresponding to $(i_{1 L|R},\ldots,i_4)$-windows, yellow square symbols for $(i_{1 L|R},\ldots,i_5)$-windows, and a green star symbols for $(i_{1 L|R},\ldots,i_6)$-windows. Let us notice that by using a $\Delta v < 10^{-6}$ multi-bounce windows with more than seven bounces will appear, but the main fractal-like features of the corresponding plot remain.

Figure~\ref{fig:F7}~(b) shows a zoom of the multi-bounce windows located after the $(1)$-window for $v>v^{R}_{(1)}$, calculated with $\Delta v = 10^{-6}$; we have numerically checked that there are no multi-bounce windows with $v<v^{L}_{(1)}$.
The red dot-dashed lines marked with a red asterisk symbol in this plot correspond to $3_R$-bounce windows, concretely $(1_{R},i_2)$-windows with $i_2= 0, 1, \ldots 94$; they clearly suggest that $\lim_{i_2 \to \infty} v_{(1_R,i_2)} = v^R_{(1)} = 0.159667$, for $i_2 \ge 0$, where a one-sided limit approaching from the right has been used, and will be used hereafter, for limits towards right boundary velocities.
The $4_R$-bounce windows marked with a blue triangle symbol in Fig.~\ref{fig:F7}~(b) correspond to twenty $(1_{R},i_2,i_3)$-windows with $i_2,i_3\ge 0$ and suggest that $\lim_{i_3\to \infty} v_{(1_R,0,i_3)} = v^L_{(1_R,0)}=0.164834$, for $i_3 \ge 0$, where a one-sided limit approaching from the left is used, and will be used hereafter, for limits towards left boundary velocities, and $\lim_{i_3\to \infty} v_{(1_R,1,i_3)} = v^L_{(1_R,1)}=0.161920$, for $i_3 \ge 0$; moreover, they could be generalized to $\lim_{i_3\to \infty} v_{(1_R,i_2,i_3)} = v^L_{(1_R,i_2)}$, for all $i_2, i_3\ge 0$.
The $5_R$-bounce windows marked with a brown filled circle symbol in Fig.~\ref{fig:F7}~(b) correspond to the $(1_{R},i_2,i_3,i_4)$-windows with $i_2,i_3,i_4\ge 0$, and suggest that $\lim_{i_4\to \infty} v_{(1_R,0,0,i_4)}\to v^R_{(1_R,0,0)}=0.164027$,
and, apparently, $\lim_{i_4\to \infty} v_{(1_R,i_2,i_3,i_4)}= v^R_{(1_R,i_2,i_3)}$, both limits for $i_2,i_3,i_4 \ge 0$.
Finally, further results shown in Fig.~\ref{fig:F7}~(b) could be generalized to $\lim_{i_{p-1}\to \infty} v_{(1_R,i_2,\ldots,i_{p-1})}= v^R_{(1_R,i_2,\ldots,i_{p-2})}$, for odd $p$, and $\lim_{i_{p-1}\to \infty} v_{(1_R,i_2,\ldots,i_{p-1})}= v^L_{(1_R,i_2,\ldots,i_{p-2})}$, for even $p$, in both cases for all $i_2,\ldots,i_{p-2}\ge 0$.

Figure~\ref{fig:F7}~(c) shows the distribution of multi-bounce windows at both sides of the $(2)$-window, for $v\in[0.1770, 0.1830]$; the plot shows $p$-bounce windows with $p=3_L$, and $4_L$ for $v<v^L_{(2)}=0.177747$, and with $p=3_R$, $4_R$, $5_R$, and $6_R$ for $v>v^R_{(2)}=0.180780$.
On the left side of the $(2)$-window in this plot, the red dot-dashed lines marked with a red asterisk symbol for $v<v^L_{(2)}$ correspond to the $3_L$-bounce windows and velocities; they clearly suggest that $\lim_{i_2 \to \infty} v_{(2_L,i_2)} = v^R_{(2)}$, for $i_2 \ge 1$; note that our numerical results do not show a $(2_L,0)$-window.
The scarce results for the $4_L$-bounce windows, marked with a blue triangle in this figure, show that $v_{(2_L,2,1)} = 0.177457 > v^R_{(2_L,2)} =  0.177443$, $v_{(2_L,4,1)} = 0.177612 > v^R_{(2_L,4)} = 0.177607$, and $v_{(2_L,6,1)} =  0.177669 > v^R_{(2_L,6)} = 0.177667$; these results suggest that $\lim_{i_3\to \infty} v_{(2_L,i_2,i_3)} = v^R_{(2_L,i_2)}$, for $i_2\ge 1$, and $i_3 \ge 0$.
On the right side of the $(2)$-window in this plot, the red dot-dashed lines marked with a red asterisk symbol for $v>v^R_{(2)}$ in Figure~\ref{fig:F7}~(c) correspond to the $3_R$-bounce windows; the velocities of the $(2_R,17)$-, \ldots, $(2_R,1)$-, $(2_R,0)$-windows clearly suggest that $\lim_{i_2 \to \infty} v_{(2_R,i_2)} = v^R_{(2)}$, for $i_2\ge 0$. The $4_R$-bounce windows marked with a  blue triangle and the corresponding values for $p=4_R$ suggest that $\lim_{i_3 \to \infty} v_{(2_R,i_2,i_3)} = v^L_{(2_R,i_2)}$, for $i_2, i_3\ge 0$.
The $5_R$-bounce windows marked with a brown filled circle in the plot suggest that $\lim_{i_4 \to \infty} v_{(2_R,i_2,i_3,i_4)} = v^R_{(2_R,i_2,i_3)}$, for $i_2, i_3, i_4\ge 0$.  The scarce results for the $6_R$-bounce windows, marked with a yellow square, show that $v_{(2_R,0,0,0,0)} = 0.182300 < v_{(2_R,0,0,0,1)} = 0.182303 <  v^L_{(2_R,0,0,0)} =  0.182306$, suggesting that $\lim_{i_5\to \infty} v_{(2_R,i_2,i_3,i_4,i_5)} = v^L_{(2_R,i_2,i_3,i_4)}$, for $i_2,i_3,i_4,i_5 \ge 0$.

Figure~\ref{fig:F7}~(d) shows the distribution of $p$-bounce windows at both sides of the $(3)$-window, for $v\in[0.1873, 0.1903]$; the plot show  multi-bounce windows with $p=3_L$ and $4_L$  for $v<v^L_{(3)}=0.187640$, and with $p=3_R$, $4_R$, and $5_R$ for $v>v^R_{(3)}=0.189220$.
On the left side of the $(3)$-window in this plot, $3_L$-bounce windows for  $v<v^L_{(3)}$ are shown in the red dot-dashed lines marked with a red asterisk symbol, clearly suggest that $\lim_{i_2 \to \infty} v_{(3_L,i_2)} = v^L_{(3)}$, for $i_2  \ge 1$; note that we have numerically searched for a $(3_L,0)$-window in our results but we have found no one.
The scarce results for the $4_L$-bounce windows, marked with a blue triangle, show that $v_{(3_L,3,1)} = 0.187555 > v^L_{(3_L,3)} =  0.187549$, and
$v_{(3_L,4,2)} = 0.187579 > v^L_{(3_L,4)} = 0.187576$;
these results suggest that $\lim_{i_3\to \infty} v_{(3_L,i_2,i_3)} = v^L_{(3_L,i_2)}$, for $i_2\ge 1$, and $i_3\ge 0$.
On the right side of the $(3)$-window in this plot, the $(3_R,29)$-, \ldots, $(3_R,1)$-, $(3_R,0)$-windows for  $v>v^R_{(3)}$ are shown in the red dot-dashed lines marked with a red asterisk symbol in Figure~\ref{fig:F7}~(d); they suggest that $\lim_{i_2 \to \infty} v_{(3_R,i_2)} = v^R_{(3)}$, for $i_2\ge 0$.
The $4_R$-bounce windows marked with a  blue triangle suggest that $\lim_{i_3 \to \infty} v_{(3_R,i_2,i_3)} = v^L_{(3_R,i_2)}$, for $i_2, i_3\ge 0$.
Finally, the $5_R$-bounce windows marked with a brown filled circle in the plot, suggest that $\lim_{i_4 \to \infty} v_{(3_R,i_2,i_3,i_4)} = v^R_{(3_R,i_2,i_3)}$, for $i_2, i_3, i_4\ge 0$.


The self-similar structure of the fractal scattering of kink-antikink solutions observed for $b=1.5$, shown in Figure~\ref{fig:F7}, is similar to that for $b=2.5$ and $3.5$; the corresponding figures are omitted here for the sake of brevity. The general features observed in our numerical experiments suggest the following conjecture about the self-similar pattern of the multi-bounce windows for the GSLeq with $b>0$. The velocity of the two-bounce windows accumulates at the critical velocity such as
$$
 \lim_{i_1 \to \infty} v_{(i_1)} = v^L_{cr} \equiv v_{cr}, \qquad i_1 \ge 1;
$$
and the velocity of the $p$-bounce windows with even $p\ge 2$ at the right side of a $2$-bounce window are
$$
 \lim_{i_{p-1}\to \infty} v_{(i_{1\,R},i_2,\ldots,i_{p-1})}
                      = v^L_{(i_{1\,R},i_2,\ldots,i_{p-2})},
 \qquad i_1 \ge 1, \ i_2,\ldots,i_{p-1}\ge 0,
$$
with even $p>2$ at the left side are as
$$
 \lim_{i_{p-1}\to \infty} v_{(i_{1\,L},i_2,\ldots,i_{p-1})}
                      = v^R_{(i_{1\,L},i_2,\ldots,i_{p-2})},
 \qquad i_1\ge 2, \ i_2 \ge 1, \ i_3,\ldots,i_{p-2}\ge 0,
$$
with odd $p>1$ at the right side of a $2$-bounce window are
$$
 \lim_{i_{p-1}\to \infty} v_{(i_{1\,R},i_2,\ldots,i_{p-1})}
                      = v^R_{(i_{1\,R},i_2,\ldots,i_{p-2})},
 \qquad i_1 \ge 1, \ i_2,\ldots,i_{p-2}\ge 0,
$$
and odd $p>1$ at the left side are as
$$
 \lim_{i_{p-1}\to \infty} v_{(i_{1\,L},i_2,\ldots,i_{p-1})}
                   = v^L_{(i_{1\,L},i_2,\ldots,i_{p-2})},
 \qquad i_1\ge 2, \ i_2 \ge 1, \ i_3,\ldots,i_{p-2}\ge 0.
$$

\subsection{Resonant energy exchange theory}
\label{sub:resonant}

\begin{figure}[t] \centering
\includegraphics[width=\textwidth/2]{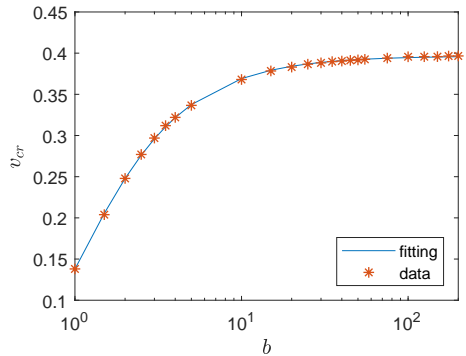}
\caption{The plot shows the critical velocity $v_{cr}$ as function of $b$ estimated from numerical calculations and fitted by Eq.~\eqref{eq:fitting:vcr}, by using $x_0=25$ for $b<25$, and $x_0=50$ for $25 \le b \le 200$; in all the simulations $\Delta t = \Delta x = 0.01$, $L=100$, and $T=500$.
}
\label{fig:F8}
\end{figure}

The results presented in the previous section for $b=1.5$ are representative of those obtained for other values of $b$. They can be understood by applying the resonant energy exchange theory developed by Campbell et al.~\cite{CampbellEtAl1983,Peyrard1983}. The critical velocity for the two-bounce windows depend on the parameter $b$; Figure~\ref{fig:F8} shows its values numerically calculated from our simulations for $b\in[1,200]$, with $x_0=25$ for $b<25$, and $x_0=50$ for $25 \le b \le 200$,  $\Delta t = \Delta x = 0.01$, $L=100$, and $T=500$. It monotonically grows as a function of $b$ apparently reaching a constant asymptotic value near $0.4$; let us guess the following ansatz
\begin{equation}
 v_{cr}(b) = \gamma\,\exp(-\alpha\,(1+1/b^\beta)),
 \label{eq:fitting:vcr}
\end{equation}
whose numerical fitting results in $\alpha = 1.056\,\pm\,0.010$, $\beta = 1.169\,\pm\,0.013$, and $\gamma = 1.14\,\pm\,0.011$ at the 95\% confidence level. Note that $v_{cr}(b)\rightarrow 0$, for $b\rightarrow 0$, as expected for the sGeq, and that the asymptotic value of the critical velocity fitted by Eq.~\eqref{eq:fitting:vcr} is $v_{cr}(\infty) = 0.3965$.  The good accuracy of this fitting is illustrated by the blue curve shown in Fig.~\ref{fig:F8}; quantitatively, the relative error is smaller than $1\%$ for $b\in[1,200]$. The analytical estimation of the critical velocity by means of using a singular perturbation theory with $\varepsilon = b$ as small parameter is outside the scope of this paper; however, let us remark that such an approach usually results in errors for $\varepsilon=1$ within $6\%$ \cite{GoodmanHaberman2004,GoodmanHaberman2005}.

\begin{figure}[t]
\includegraphics[width=\textwidth ]{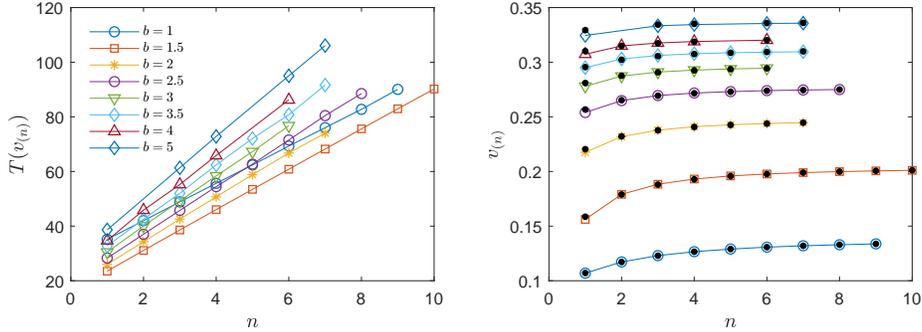}
\caption{Left plot shows the width in time of the $(n)$-window centered in $v_{(n)}$ as a function of the number of oscillations between  the  bounces for $b=1$, 1.5, 2, 2.5, 3, 3.5, 4, and 5; right plot shows $v_{(n)}$ (coloured continuous line) compared with its estimation by Eq.~\eqref{eq:central:velocity} (black dots)  for $b=1$, 1.5, 2, 2.5, 3, 3.5, 4, and 5; in all the simulations $x_0=25$, $\Delta t = \Delta x = 0.01$, $L=100$, and $T=500$.
}
\label{fig:F9}
\end{figure}

The resonant energy exchange theory~\cite{CampbellEtAl1983,Peyrard1983,Manton2021} explains the 2-bounce windows in the interaction of the kink and the antikink as the result of the excitation of internal modes  during the collision. Assuming a single internal mode with frequency $\omega_B$, the time between the first and second bounces $T(v_{(n)})$, i.e., the width in time of the oscillatory waveform shown in Fig.~\ref{fig:F5} at $u(0,t)=0$ in the $b=1.5$ case, satisfy the resonance condition
\begin{equation}
 w_B\,T(v_{(n)}) = 2\pi\,n+\delta,
 \label{eq:resonance:condition}
\end{equation}
for the $(n)$-window, where $\delta\in[0,2\pi)$  is an offset phase (to be numerically fitted). The value of $T({v_{(n)}})$ can be easily estimated by using the inverse linear interpolation solution of equation $u(0,t;v_{(n)})=0$. Figure~\ref{fig:F9} (left plot) shows that $T({v_{(n)}})$ is a linear function of $n$ for $b=1$, 1.5, 2, 2.5, 3, 3.5, 4, and~5; in all the simulations $x_0=25$, $\Delta t = \Delta x = 0.01$, $L=100$, and $T=500$.
Table~\ref{tab:resonances} shows the values of $\omega_B$ and $\delta$ with 95\% confidence intervals after numerical fitting of Eq.~\eqref{eq:resonance:condition} with the data plotted in Fig.~\ref{fig:F9} (left plot); the internal mode frequency $\omega_B$ decreases as $b$ increases.

\begin{table}
\centering
\scriptsize
\caption{Coefficients $\delta$, $\omega_B$, and $\beta$ from the numerical fitting of Eqs.~\eqref{eq:resonance:condition} and~\eqref{eq:width:time} by using the data shown in the left plot of Fig.~\ref{fig:F9}  with 95\% confidence intervals, and $\omega_S$ for $v=0$ calculated by solving the eigenvalue problem~\eqref{martinvergara-Schrodinguer}; the last column shows the percentage error between $\omega_B$ and $\omega_S$.
}

\begin{tabular}{cccccc}
\hline
$b$   & $\delta$ & $\omega_B$ & $\beta$ & $\omega_S$ & $\omega_B\sim\omega_S$  \\
$1$   & $26.04\,\pm\,0.38$ & $0.9197\,\pm\,0.0059$  &  $3.0599\,\pm\,0.0027$ & $0.9396$ & $2.2\%$\\
$1.5$ & $13.84\,\pm\,0.22$ & $0.8483\,\pm\,0.0035$ & $3.0409\,\pm\,0.0221$   & $0.8688$ & $2.4\%$\\
$2$   & $14.26\,\pm\,0.91$ & $0.7809\,\pm\,0.0173$ & $2.9848\,\pm\,0.0477$   & $0.8020$ & $2.7\%$\\
$2.5$ & $14.41\,\pm\, 0.40$ & $0.7286\,\pm\,0.0065$ & $2.9552\,\pm\,0.0750$  & $0.7446$ & $2.2\%$\\
$3$   & $14.67\,\pm\,0.54$ & $0.6828\,\pm\,0.0095$ & $2.9437\,\pm\,0.1263$   & $0.6961$ & $1.9\%$\\
$3.5$ & $14.80\,\pm\,0.92$ & $0.6438\,\pm\,0.0142$ & $3.2802\,\pm\,0.1456$   & $0.6550$ & $1.7\%$ \\
$4$   & $15.19\,\pm\,1.24$ & $0.6131\,\pm\,0.0206$ & $3.0656\,\pm\,0.1779$   & $0.6170$ & $0.6\%$\\
$5$   & $15.42\,\pm\,0.57$ & $0.5590\,\pm\,0.0071$ &  $2.8531\,\pm\,0.3556$  & $0.5631$ & $0.7\%$\\\\
\end{tabular}
\label{tab:resonances}
\end{table}

Moreover, assuming that the energy ``stored'' in the internal mode is roughly conserved, the resonant energy exchange theory~\cite{CampbellEtAl1983,Peyrard1983} predicts the time width
\begin{equation}
 T(v_{(n)}) = \frac{\beta}{\sqrt{v_{cr}^2-v_{(n)}^2}},
 \label{eq:width:time}
\end{equation}
where $\beta$ is an empirically determined constant. Table~\ref{tab:resonances} shows the values of $\beta$ with 95\% confidence intervals after numerical fitting of Eq.~\eqref{eq:width:time} with the data plotted in Fig.~\ref{fig:F9} (left plot). Combining Eqs.~\eqref{eq:resonance:condition} and~\eqref{eq:width:time}, the central velocity of the $(n)$-window can be estimated as
\begin{equation}
 v_{(n)} \approx \sqrt{ v_{cr}^2 - \frac{\beta^2\,\omega_B^2}{(2\pi\,n+\delta)^2} }
 .
 \label{eq:central:velocity}
\end{equation}
Hence the values of $v_{(n)}$, $v_{(n)}^L = v_{(n)}-T(v_{(n)})/2$, and $v_{(n)}^R = v_{(n)}+T(v_{(n)})/2$ can be easily estimated for $(n)$-windows. Figure~\ref{fig:F9} (right plot) compares Eq.~\eqref{eq:central:velocity} (black dots) with  $v_{(n)}$ (coloured continuous line) showing the good accuracy of the approximation as $n$ increases.

\begin{table}
\centering
\scriptsize
\caption{Comparison of the theoretical estimation $\beta_{th}$ given by Eq.~\eqref{eq:beta:th} with $\beta$ in Table~\ref{tab:resonances}, including the percentage of the error between both values, for $b=1.5$, $2.5$, and $3.5$. The parameter $\alpha$ in Eq.~\eqref{eq:beta:th} was estimated by using $v\in[v_1,v_2]$ with $\Delta v = 0.001$.
}
\begin{tabular}{ccccccr}
\hline
$b$   & $v_{cr}$ & $[v_1,v_2]$ & $\alpha$ & $\beta$ & $\beta_{th}$ & $\beta\sim\beta_{th}$ \\
$1.5$ & 0.204 & $[0.205, 0.210]$ & $1.235\pm 0.028$ & $3.04\pm 0.02$ & $ 2.827\pm 0.032$ & $7\pm 2\,\%$\\
$2.5$ & 0.277 & $[0.280, 0.310]$ & $1.521\pm 0.027$ & $2.96\pm 0.08$ & $2.547\pm 0.022$ & $14\pm 4\,\%$\\
$3.5$ & 0.312 & $[0.320, 0.355]$ & $1.657\pm 0.015$ & $3.28\pm 0.15$ & $2.441\pm 0.011$ & $26\pm 6\,\%$
\end{tabular}
\label{tab:resp}
\end{table}

Campbell et al.~\cite{CampbellEtAl1986} presents a theoretical estimation $\beta_{th}$ of the empirical parameter $\beta$ given by
\begin{equation}
\beta_{th} = \frac{\pi}{\omega_0\,\sqrt{\alpha}},
 \label{eq:beta:th}
\end{equation}
where $\omega_0^2 =
d^2G(u_0)/du^2 = 1$, for $dG(u_0)/du=0$ (note that $u_0$ depends on $b$, but $\omega_0$ is independent of $b$ for the GSLeq) and $\alpha$ is another empirical parameter determined by fitting $v_{out}^2 =\alpha\,(v^2-v_{cr}^2)$, for $v>v_{cr}$, i.e., the black curve at the left part of the plot in Fig.~\ref{fig:F7}~(a). Table~\ref{tab:resp}
compares the theoretical estimation~\eqref{eq:beta:th} with the estimation of $\beta$ in Table~\ref{tab:resonances}, showing that $\beta_{th}$ underestimates $\beta$; moreover, the percentage of the error between both values increases as $b$ does.

\begin{figure}[t] \centering
\includegraphics[width=\textwidth]{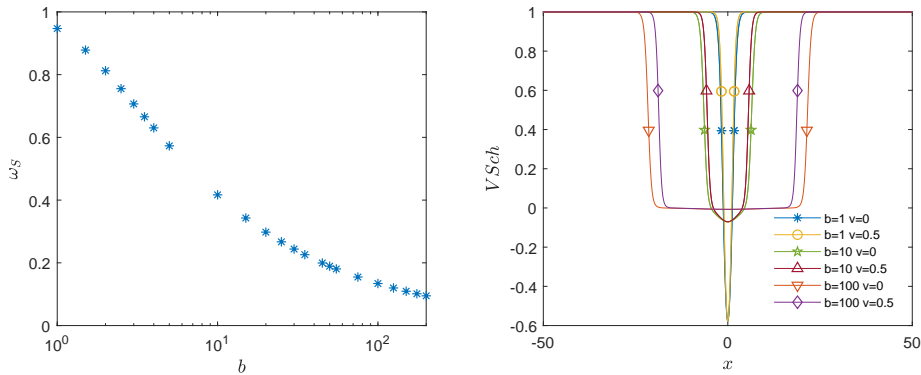}
\caption{The left plot shows the mode frequency $\omega_s$ numerically calculated by solving the eigenvalue problem for the Schr\"odinger equation~\eqref{martinvergara-Schrodinguer}, by using $\Delta x = 0.01$, $L=100$, and $v=0.2$. The right plot shows the potential $V_{Sch}(x)$ for $b=1,10$, and $100$, with $v=0$, and $0.5$.
}
\label{fig:F10}
\end{figure}

In order to analyze the small oscillations about the kink waveform in the multi-bounce windows, Campbell et al.~\cite{CampbellEtAl1983} introduced a Schr\"odinger equation for small perturbations of the 0-kink given by Eq.~\eqref{martinvergara-kkeq:static:solution}. The introduction of $u_{k,0}(x)+\delta u(x,t)$ into Eq.~\eqref{martinvergara-kkeq} results in
$$
 \ndpar{\delta u}{t}{2} - \ndpar{\delta u}{x}{2}
 + \ndtot{G(u_{k,0})}{u}{2}\, \delta u = 0,
$$
whose  Fourier transformation in time, $\delta u(x,t) = \delta \hat{u}(x)\,\exp(-\mbox{i}\,\omega_S\,t)$, yields
\begin{equation}
  -\ndtot{\delta\hat{u}}{x}{2} + V_{Sch}(x)\,\delta\hat{u} = \omega_S^2\,\delta\hat{u},
  \label{martinvergara-Schrodinguer}
\end{equation}
where
$$
 V_{Sch}(x) = \frac{4\,(1+b^2)\,\cos(u_{k,0})-b^2\,(3+\cos(2\,u_{k,0}))}
                   {4\,(1+b^2\,(1-\cos(u_{k,0})))^{3/2}},
$$
which can be easily evaluated once the kink is obtained by using a numerical quadrature formula. Figure~\ref{fig:F10} (right plot) shows the potential $V_{Sch}(x)$ for $b=1, 10$, and $100$, and two velocities $v=0$, and $0.5$. The potential has only one well, instead of the double well observed for the double sine-Gordon equation in Ref~\cite{CampbellEtAl1986, GaniEtAl2018}; the sharp-pointed shape of the potential for $b=\OL{1}$, with a negative minimum, widens to a square-like shape for $b\gg 1$, as the minimum approaches to $0^-$. Hence, the Schr\"odinger equation~\eqref{martinvergara-Schrodinguer} has ``Goldstone" mode with $\omega_S=0$ and a continuous spectrum for $\omega_S\ge 1$; additionally there are several trapping modes, whose number increases as the potential widens as $b$ grows up.

The application of the finite difference operator~\eqref{eq:fdm4} to Eq.~\eqref{martinvergara-Schrodinguer} results in an eigenvalue problem which allow the determination of $\omega_S^2$. Figure~\ref{fig:F10} (left plot) shows $\omega_S$ for the first excited mode as a function of $b$; the frequency monotonically decreases approaching the ground state $\omega_S=0$ as $b$ increases. Table~\ref{tab:resonances} shows that $\omega_S$ overestimates $\omega_B$ with an error smaller than $3\%$.

\section{Conclusions}
\label{sec:conclusions}

The kink-antikink collision of the graphene superlattice equation, introduced by Kryuchkov  and Kukhar’ to study the nonlinear propagation of electromagnetic waves in Ratnikov’s graphene superlattice, is studied numerically for the first time. The most efficient method in the comparison of implicit Pad\'e numerical schemes for the sine-Gordon equation in Ref.~\cite{MartinVergaraEtAl2019} has been selected; concretely, a fourth-order in space generalization of Strauss--V\'azquez method, which is the second-order in time and uses a treatment of the nonlinearity that ensures good energy conservation. The collision of a kink and an antikink with the same, but opposite speed is studied under periodic boundary conditions, so an imbricated initial condition is used.

A new multi-index notation to formalize the quasi-fractal pattern observed in the multi-bounce windows with arbitrary number of bounces is introduced. For speeds below a critical velocity the collision results in the formation of an oscillatory bound state, resembling a pseudo-breather, with slowly diminishing amplitude as time increases; such a behaviour is general, except in a series of windows where the solution bounces several times before the kink and the antikink escape to infinity. The multi-bounce windows show a quasi-fractal structure depending on the number of bounces.

The series of two-bounce windows approach from the left to the critical velocity, that acts as a limit point; in the process the two-bounce windows reduce its width and increase the number of oscillations before escaping. {A series} of three-bounce windows accumulates at both sides of the two-bounce windows which boundaries acting as the corresponding critical velocities; note that, differing from the result for the $\phi^4$ model~\cite{GoodmanHaberman2005}, there is no three-bounce windows in the left side of the first two-bounce window, but only in the right side. Four-bounce windows accumulates to the (right-) left-side of the three-bounce windows located at the (left-) right-side of the two-bounce windows. This general pattern repeats for multi-bounce windows with a larger number of bounces.

In order to understand the quasi-fractal structure observed in the scattering of kinks and antikinks, the resonant energy exchange theory developed by Campbell and collaborators is applied. The time between the first and second bounces in the two-bounce windows is a linear function of the number of oscillations before escaping; its numerical fitting allows the determination of the central velocity of the two-bounce windows with a good accuracy. The frequency of the resonance mode responsible for the trapping of the kink and the antikink inside the two-bounce windows can be estimated by solving an eigenvalue problem for a Schr\"odinger equation with an effective potential; this approximation is better as the geometrical parameter of the graphene superlattice equation increases, the potential widens and the frequency monotonically decreases.

Further work is certainly required to apply the variational approximation developed by Goodman and Haberman~\cite{GoodmanHaberman2007,Goodman2008} for the graphene superlattice equation with small $b$; the reduced variational ordinary differential model introduced by these authors should illuminate the multi-index notation introduced in this paper for the observed fractal-like window patterns. Moreover, the application of perturbation methods based on the inverse scattering transform for the sine-Gordon equation to the graphene superlattice equation with small geometrical parameter is expected to yield accurate analytical approximations that could complement the results of the resonant energy exchange {theory~\cite{KivsharMalomed1989}.
In particular, radiation losses makes possible that fusion of a kink-antikink pair results into a breather-like solution;  a fusion threshold was identified in Ref.~\cite{Malomed1985} for the double sine-Gordon and one is expected for the graphene superlattice equation. In fact, the study of pseudo-stable breather-like solutions of the graphene superlattice equation requires deep study.}
Another future line of research is the exploration of the spectral walls~\cite{AdamEtAl2019,AdamEtAl2019b,AdamEtAl2020b,AdamEtAl2020c} surrounding the solitons of the graphene superlattice equation and its role in the scattering of kinks and antikinks.
Finally, the kink-antikink collisions of the integro-differential equation introduced by  Kryuchkov, Kukhar’ and Zav’yalov~\cite{KryuchkovEtAl2013} for the modelling of the electromagnetic waves in the graphene superlattice under a high-frequency field deserves attention.

\section*{Acknowledgements}
{The authors thank the reviewers for their thoughtful comments and efforts towards improving our manuscript.} The research reported here was partially supported by Projects DeepBIO (TIN2017-85727-C4-1-P) of the Programa Estatal de Fomento de la Investigaci\'on Cient\'ifica y T\'ecnica de Excelencia del Ministerio de Ciencia e Innovaci\'on of Spain, and RoCoSoyCo (UMA18-FEDERJA-248) of the Consejer\'ia de Econom\'ia y Conocimiento, Junta de Andaluc\'ia, Spain.

\bibliographystyle{elsarticle-num}

\end{document}